\documentclass[onecolumn]{aastex631}

\usepackage{multirow,colortbl,chemformula,booktabs,graphicx,amsmath}

\usepackage{rotating}
\usepackage{graphicx,journal_names}
\newcommand{\RHK}{R$^\prime_\mathrm{HK}$}

\newcommand{\lmm}{L$_{\rm 95GHz}$}

\newcommand{\Msun}{M$_\odot$}

\newcommand{\df}{$\delta$}
\newcommand{\Snu}{$\mathrm{S(\nu)}$}

\newcommand{\Prot}{${\rm P_{rot}}$}

\newcommand{\Smod}{$\mathrm{S^{ph}_{mod}(\nu)}$}
\newcommand{\Smodc}{$\mathrm{S^{ch}_{mod}(\nu)}$}
\newcommand{\dsmod}{$\mathrm{\Delta S/S^{ph}_{mod}}$}
\newcommand{\dsmodc}{$\mathrm{\Delta S/S^{ch}_{mod}}$}

\newcommand{\teff}{T$_\mathrm{eff}$}
\newcommand{\Tbp}{T$_\mathrm{B}(\nu)$}

\newcommand{\alfmm}{$\mathrm{\alpha_{mm}}$}
\accepted{June 20, 2025}

\submitjournal{ApJ}

\begin{document}

   %\title{ }
    %\title{Exploring quiescent chromospheric heating and activity in AD\,Leo using millimeter observations} 
%\title{Importance of non-thermal emission processes in modeling the millimeter spectrum of a young M-dwarf even during quiescence.}
\title{Role of non-thermal processes in the quiescent and active millimeter spectrum of a young M-dwarf.}
%    \titlerunning{Origin of quiescent and active mm emission in a young dM: Role of corona}
 \correspondingauthor{Atul Mohan}
\email{atul.mohan@nasa.gov}

\author[0000-0002-1571-7931]{Atul Mohan}
\affiliation{NASA Goddard Space Flight Center, Greenbelt, MD 20771, USA}
\affiliation{The Catholic University of America, Washington, DC 20064, USA}

\author[0000-0002-8077-5572]{Peter H. Hauschildt}
\affiliation{Hamburger Sternwarte, Gojenbergsweg 112, 21029 Hamburg, Germany}

\author[0000-0001-8321-5514]{Birgit Fuhrmeister}
\affiliation{Hamburger Sternwarte, Gojenbergsweg 112, 21029 Hamburg, Germany}
\affiliation{Th\"uringer Landessternwarte Tautenburg, Sternwarte 5, D-07778 Tautenburg, Germany}

\author[0000-0002-2325-5298]{Surajit Mondal}
\affiliation{Center for Solar-Terrestrial Research, New Jersey Institute of Technology, 323 M L King Jr Boulevard, NJ 07102-1982, USA}

\author[0000-0003-4452-0588]{Vladimir Airapetian}
\affiliation{NASA Goddard Space Flight Center, Greenbelt, MD 20771, USA}
\affiliation{American University, 4400 Massachusetts Avenue NW, Washington, DC 20016 USA}

\author[0000-0002-5006-7540]{Sven Wedemeyer}
\affiliation{Rosseland Centre for Solar Physics, University of Oslo, Postboks 1029 Blindern, N-0315 Oslo, Norway}
\affiliation{Institute of Theoretical Astrophysics, University of Oslo, Postboks 1029 Blindern, N-0315 Oslo, Norway}
   

%   \institute{Rosseland Centre for Solar Physics, Institute of Theoretical Astrophysics - University of Oslo, Blindern 0315, Oslo\\
         %\and
          %   University of Alexandria, Department of Geography, ...\\
             
             %\thanks{The university of heaven temporarily does not accept e-mails}
             
%\authorrunning{Mohan, A. et al.}

%  \date{Received September 15, 1996; accepted March 16, 1997}
 %  \date{Received  - ; accepted  - }

% \abstract{}{}{}{}{} 
% 5 {} token are mandatory
 
  \begin{abstract}
  %Why are we doing this? Background\\
 { Millimeter (mm) emission from F - M dwarfs (cool stars) primarily traces chromospheric activity, with thermal emission thought to dominate in quiescence. Despite the high chromospheric activity, the quiescent mm spectral fluence (mm-\Snu) of young ($<$1\,Gyr) M dwarfs (dMs) remain largely unexplored.
We present the quiescent mm-\Snu\ of a young dM, AD\,Leo, observed around 94\,GHz using the Northern Extended Millimetre Array (NOEMA).
The observed quiescent mm-\Snu\ exceeds the thermal flux density from a 1D chromospheric model, constrained by optical-UV spectroscopic data, by up to a factor of 7. This indicates a quasi-steady non-thermal emission powered by supra-thermal electrons unlike in old ($>$1\,Gyr) cool stars
, whose quiescent mm-\Snu\ generally agree with 1D thermal models. 
The mm-brightness temperature spectral index (\alfmm; \Tbp$\sim\nu^{\rm -\alpha_{mm}}$) of AD\,Leo deviates by a factor of 3 from the \alfmm–\teff\ scaling law for old sun-like stars~\citep{Atul22_EMISSAII}, while UV\,Ceti, an older M6V star, follows the trend.
Also, we report a double-hump flare with second-scale variability in flux density and spectral index, and a frequency-rising nature with brightness increasing with frequency.
The flare resemble certain solar events, but is unlike the second-scale events reported in dMs. The non-thermal flare humps suggest multiple injections of accelerated electrons. The mean flare luminosity ($\sim 2 - 5 \times10^{15}$\,erg\,s$^{-1}$\,Hz$^{-1}$) and duration ($18\pm 2$\,s) are comparable to flares reported in AU\,Mic and Proxima\,Cen, but 100 - 1000 times weaker than the minutes-long dM flares observed by the South Pole Telescope.
}
\end{abstract}

%   \keywords{stars: flare - stars: activity - stars: magnetic field - stars: coronae - radio continuum: stars - stars: low-mass

 %  \maketitle
%
%-------------------------------------------------------------------

%=============================================================================
%=============================================================================
%=============================================================================
\section{Introduction}
Low-mass main-sequence stars (M-dwarfs; dMs) host most of the known Earth-like exoplanets residing in habitable zones, making them interesting targets for exploring the active outer atmospheric structure and the emergence of activity that can impact exoplanetary habitability~\citep{Bashi20_occurrence_of_smallplanetsFGK, vladimir20_SpW_impact_exoplanets}
Emission at millimeter (mm) frequencies is expected primarily to trace the chromospheric thermal emission across heights in cool main-sequence stars (Spectral type: F - M)~\citep{Sven16_ALMA_science,White20_MESAS,Atul21_EMISSAI,Atul23_mmRadioTomo}. However, during flares, the mm spectrum can have non-thermal contributions from flare-accelerated electrons in active regions, providing insights on the local particle acceleration and magnetic field strengths~\citep{Macgregor20_AUmicflare}. 
Detecting the quiescent mm emission of nearby stars remained difficult until recently due to the lack of sensitive instruments. Even with the advent of modern telescopes like ALMA and NOEMA, the data on main sequence stars are still too sparse to make any robust statistical inferences or constrain stellar chromospheric models for different spectral types~\citep{Macgregor20_AUmicflare,Atul21_EMISSAI}.
\cite{Atul21_EMISSAI} compiled all robust detections of quiescent mm emission from main-sequence stars in the ALMA science archive and explored chromospheric heating across the stellar main sequence.
Using this sample, \cite{Atul22_EMISSAII} derived the mm-brightness temperature spectrum (\Tbp) for each star and estimated its spectral index (\Tbp$\propto \nu^{-{\rm \alpha_{mm}}}$;\alfmm). 
The authors explored the relationship between \alfmm\ and stellar physical parameters,  especially the effective surface temperature \teff, demonstrating how \alfmm\ traces the atmospheric heating gradient across the main-sequence. 
{ Their analysis suggested that \alfmm\ can be a meaningful metric of steady chromospheric heating and activity.} 

Though studies in mm band are sparse, the nature of quiescent chromospheric emission and its trends with physical parameters have been explored over decades using high energy spectroscopic metrics~\citep[e.g.,][]{Wilson68_Sindex,Noyes84_RHK, Herbig85_RHalph_defn_linktoRHK, stepien94_Defn_Rx+Ro_Vs_activity_n_manyCorCurves,Vidotto14_B_Vs_age_n_rot,2014MNRAS.444.3517M,Linsky16_Stellar_chromRev,2023A&A...673A.137P,2024MNRAS.535.2394F}.
Based on the trends shown by the common chromospheric activity metrics \RHK\,(Ratio of Ca-II H \& K flux to the bolometric flux), \cite{Barnes03_Rot_Vs_age_Vs_Activity} showed that the stars fall under two different activity groups, namely `C' and `I', in the rotational period (\Prot) - age - \teff\ plane. The C branch consists of young ($<$1\,Gyr) fast rotators (\Prot$\lesssim$5\,d), while the I branch comprises old ($>$1\,Gyr) slow-rotating stars.
Stars tend to migrate from the active C to the I branch over time as they spin down due to magnetic braking and wind-driven mass loss~\citep[e.g.,][]{vanSaders16_P-a-teff_magbreak,2018ApJ...862...90G}.
Hence, to understand the nature of chromospheric structure and emergent activity, it is important to sample both populations across the \Prot-age-\teff\ plane.
{ The study by \cite{Atul22_EMISSAII} was restricted to a small sample of F - K type I branch stars, for all of which the mm-emission was demonstrated by various authors~\citep[e.g.][]{Sven16_ALMA_science,2018MNRAS.481..217T,White20_MESAS,Atul21_EMISSAI} to be consistent with chromospheric thermal emission using data-constrained 1D models.
However, there has been no systematic study on a sample of active C branch stars due to lack of clear detections of quiescent stellar mm-emission across frequencies. A major technical hurdle has also been contamination from unresolved companions and circumstellar disks, both of which can have significant mm emission~\citep[e.g.][]{2015A&A...576A..72L,2017MNRAS.465.2595M,2017ApJ...842....8M,booth17_epsEriALMA,Long19_ALMAdisksurveyTaurus,ansdell20_youngDpperstarDisk,Macgregor20_AUmicflare,damian23_YSOdisks_Orioncluster}.
In the sample of \citep{Atul21_EMISSAI}, the authors had systematically removed candidate C branch stars with known unresolved disks based on IR and optical survey data.
The work presented here is part of an ongoing effort to sample F–M dwarfs across the \Prot–age–\teff\ plane and systematically characterize their chromospheric quiescent and flare emission, as probed by mm-\Tbp, by choosing stars with no known companions or disks within the telescope resolution limit.}

{ This study addresses AD\,Leo, a young ($\sim$250\,Myr) M3.5V star close to the fully convective internal core transition boundary owing to its mass~\citep[0.4\Msun;][]{2021A&A...645A.100S} and \teff~\citep[$\sim$3477\,K; ][]{Soubiran16_Ref3} ~\citep{2008MNRAS.390..567M,2015ApJ...804...64M,Chabrier00_lowmass_n_BD_struc,Donati09_Rev_Bfield}. 
Located at a distance of $\sim 5$\,pc~\citep{GaiaDR32021}, the star shows a high optical flaring rate $\approx$ 0.6 h$^{-1}$~\citep{dal2020flare}. }
Studies that monitored the star in spectral lines and X-ray continuum suggest high levels of chromospheric and coronal activity~\citep{vandenBesselaar03_ADLeoXMMflares,Robrade05_ADLeoXrayprops,crespo06_ADLeo_Spectrallineflares}. 
Recently, \cite{Atul24_ADleotypeIV} reported the signatures of type-IV and probable type-III burst features in the radio dynamic spectrum of AD\,Leo, presenting the first detection of {radio burst types often associated with solar coronal mass ejection events in a non-solar type young active dM.}
Hence, AD\,Leo is a compelling candidate to probe the mm-\Tbp, contrast with the existing atmospheric models, and explore its upper atmospheric structure.

Section~\ref{sec:data} describes the data and analysis, while Sec.~\ref{sec:results} will present the results. 
The analysis of the results and inferences will be discussed in Sec.~\ref{sec:disc}. Section~\ref{sec:conclusion} will present our conclusions.
\section{Data and Methodology}\label{sec:data}
% \usepackage{multirow}
% \usepackage{graphicx}
% \usepackage[normalem]{ulem}
% \usepackage{multirow}
% \usepackage{graphicx}
% \usepackage[normalem]{ulem}
\subsection{Observations}
% \usepackage{multirow}
% \usepackage{graphicx}
% \usepackage[normalem]{ulem}
\begin{table*}[]
\centering
\resizebox{\textwidth}{!}{%
\begin{tabular}{|l|lc|c|c|}
\hline
\multirow{2}{*}{\textbf{Date}} & \multicolumn{2}{c|}{\textbf{Calibrators}} & \multirow{2}{*}{\textbf{Comments}} & \multirow{2}{*}{\textbf{\begin{tabular}[c]{@{}l@{}}\hspace{-0.85cm}On-source\\ \hspace{-0.4cm}time\end{tabular}}} \\ \cline{2-3}
 & \multicolumn{1}{l|}{\textbf{Bandpass}} & \textbf{Phase} &  &  \\ \hline
31-May & \multicolumn{1}{l|}{3C273} & 0953+254{\it,J1012+232} & \begin{tabular}[c]{@{}c@{}}Good calibration solutions were obtained. \\ Only 4\% data flagged.\end{tabular} & 2h 48m \\ \hline
3-Jun & \multicolumn{1}{l|}{0851+202} &\hspace{-0.79cm}\begin{tabular}[c]{@{}c@{}}{\it MWC349}, 0953+254\\ \hspace{-0.1cm}{\it J1012+232, 2010+723}\end{tabular} & \begin{tabular}[c]{@{}c@{}}Great weather conditions. Excellent calibration \\ solutions. 0\% data flagged.\end{tabular} & 2h 12m \\ \hline
6-Jun & \multicolumn{1}{l|}{3C279} &\hspace{-0.7cm}\begin{tabular}[c]{@{}c@{}}{\it MWC349, J1012+232},\\\hspace{-0.1cm} {\it 2010+723}\end{tabular} & \begin{tabular}[c]{@{}c@{}}Great weather conditions. Excellent calibration \\ solutions. 0\% data flagged.\end{tabular} & 1h 54m \\ \hline
14-Jul & \multicolumn{1}{l|}{3C84} & {\it LKHA101, J1012+232}& \begin{tabular}[c]{@{}c@{}}Occasional bad weather. Good amplitude calibration \\ but phase calibration was affected. 61\% data flagged.\end{tabular} & 4h 30m \\ \hline
19-Jul & \multicolumn{1}{l|}{3C273} & {\it J1012+232} & Occasional bad weather. 61\%data flagged. & 1h 6m \\ \hline
7-Nov & \multicolumn{1}{l|}{0851+202} & {\it LKHA101, J1012+232} & Good weather overall. 0\% data flagged. & 1h 6m \\ \hline
\end{tabular}%
}
\caption{Details of data calibration and total on-source time for various observing sessions. Bandpass and phase calibrators
are mentioned. The phase calibrators with an archived model spectrum are marked in italics.}
\label{tab:calib}
\vspace{-0.4cm}
\end{table*}
AD\,Leo was observed with the Northern Extended Millimeter Array~\citep[NOEMA;][]{Chenu16_NOEMA_techpaper} in Band 1 centered around 94\,GHz as a part of the observing proposal W22BF (PI: Atul Mohan). 
{Band 1 has two sidebands (SBs)}, namely lower (LSB: 82 - 90\,GHz)  and upper (USB: 98 - 106\,GHz)
Interferometric observations were performed using ten 15-m telescopes on several observing sessions between May 31, 2023 and Nov 7, 2023. The total on-source time is around 13.5\,h in the standard `Line' (L) and 
`Continuum' (C) { observing modes across two SBs.
LSB and USB are further divided into 4\,GHz wide basebands\footnote{\href{https://www.iram.fr/IRAMFR/GILDAS/doc/html/noema-intro-html/node6.html}{https://www.iram.fr/IRAMFR/GILDAS/doc/html/noema-intro-html/node6.html}.}}
The L mode produced stellar spectra at 45\,s and 2\,MHz averaging, while the C mode generated 4\,GHz and 1\,s averaged data in LSB and USB.
The L mode is ideal for detecting a high-significance mean quiescent flux spectrum, which is expected to be weak given the instrumental sensitivity, with required spectral averaging. Meanwhile, the C mode is ideal to study mm flares, which are expected to be orders of magnitude brighter than the weak quiescent emission and vary at second-scale in active dMs~\citep{MacGregor18_proxima_flares, Macgregor20_AUmicflare}.
The observations were made across two orthogonal linear polarization axes, horizontal (H) and vertical (V).
The L and C  data along the two polarizations were calibrated using the standard NOEMA pipeline, Continuum and Line Interferometer Calibration (CLIC) software\footnote{\href{https://www.iram.fr/IRAMFR/GILDAS/doc/html/clic-html/clic.html}{https://www.iram.fr/IRAMFR/GILDAS/doc/html/clic-html/clic.html}}, with the help of a NOEMA-assigned support astronomer.
The details of the various observation sessions, weather conditions, calibrators, and quality of the calibration solutions are summarized in Table~\ref{tab:calib}. 
For each session, there was a bandpass calibrator and a set of commonly used `standard' phase calibrators of which at least one has a regularly updated model spectrum around 100 GHz in the {NOEMA calibrator archive}. After the antenna-wise bandpass calibration, CLIC software computes the fluxes and spectral indices of the phase calibrators by applying the bandpass solutions. 
We discarded the data from periods when the calibrated spectrum of the standard phase calibrators did not agree with the model in the NOEMA archive.
In the remaining calibrated data, we detected a flare on 14 July 2023.
The calibrated flare data in the two polarizations were separated from the quiescent emission data, and the resultant data sets were written out as separate measurement sets, readable by the Common Astronomy Software Applications~\citep[CASA;][]{CASA2022} software. 

\subsubsection{Quiescent spectrum}\label{sec:Q_data}
The CASA task \texttt{tclean} was used to image the calibrated {L mode data} with varying time-frequency averaging in Stokes I. 
{The images across the spectro-temporal axes in which a phase center source is detected at $\gtrsim 5 \sigma$ significance, where $\sigma$ is the root mean square noise level in the black sky region, were chosen.
The following conditions were imposed on every chosen image to ensure that the detection is reliable, despite imaging artifacts.
\begin{enumerate}
    \item Gaussianity of the blank sky noise pixel flux density distribution.
    \item No negative flux pixels with absolute flux $>50$\% of the phase center source are present in the field of view.
    \item The presence of the deepest negative point is justified given the blank sky noise statistics and the number of pixels in the image.
\end{enumerate}
The first condition ensures that the $5\sigma$ thresholding is meaningful given the true image plane noise statistics.
Imposing these conditions led to the further discarding of images, despite a $>5\sigma$ phase center source detection. 
The point source flux densities of the final set of reliable stellar detections were estimated using the standard CASA task \texttt{imfit} with the built-in noise offset subtraction feature.}

To explore the star's full quiescent spectral energy distribution (SED), we need data beyond the mm band. The Strasbourg Data Centre\footnote{\href{http://vizier.unistra.fr/vizier/sed/}{http://vizier.unistra.fr/vizier/sed/}} (CDS) was queried to gather the archival stellar spectral flux density data in the THz to ultraviolet (UV), i.e., $10^3$ - $10^6$\,GHz, range. The detailed procedure of querying and flagging bad data points to obtain a reliable SED is elaborated in \cite{Atul21_EMISSAI}.
The SED data (hereafter, \Snu) was compared with {model SEDs, \Smod\ and \Smodc. \Smod\ tracks purely photospheric emission, while \Smodc\ is based on a model with a photosphere and chromosphere constrained by the Ultraviolet and Visual Echelle Spectrograph \citep[UVES;][]{dekker00_uves} spectroscopic data. Refer Sec.~\ref{sec:dis_quiescent} for details.}
%those derived from the following atmospheric models.}

\subsection{The millimeter flare}
%{The observations on 14 July 2023 detected a flare with flux density ranging from 25 to 110,mJy, as opposed to the quiescent flux density level ($\sim$0.09 - 0.15,mJy).
Though observation in the preflare period was affected by intermittent bad weather, leading to flagging of data (see Table~\ref{tab:calib}), the flare period, which lasted only for $\sim$\,24\,s, was unaffected. The flare data in the C mode were imaged using the CASA task \texttt{tclean}, with varying time–frequency averaging. The images were then self-calibrated in amplitude and phase to further correct for residual antenna gain calibration errors. For further analysis, reliable images were selected that satisfied the conditions elaborated in Section~\ref{sec:Q_data}. We selected band-averaged LSB and USB images with 1\,s temporal averaging to study Stokes I flux and spectral index evolution, while those with 2\,s time averaging in the H and V polarizations to explore polarization variability.

\section{Results}\label{sec:results}
\begin{figure*}[!htb]
\centering  
\includegraphics[width=\textwidth,height=0.5\textheight]{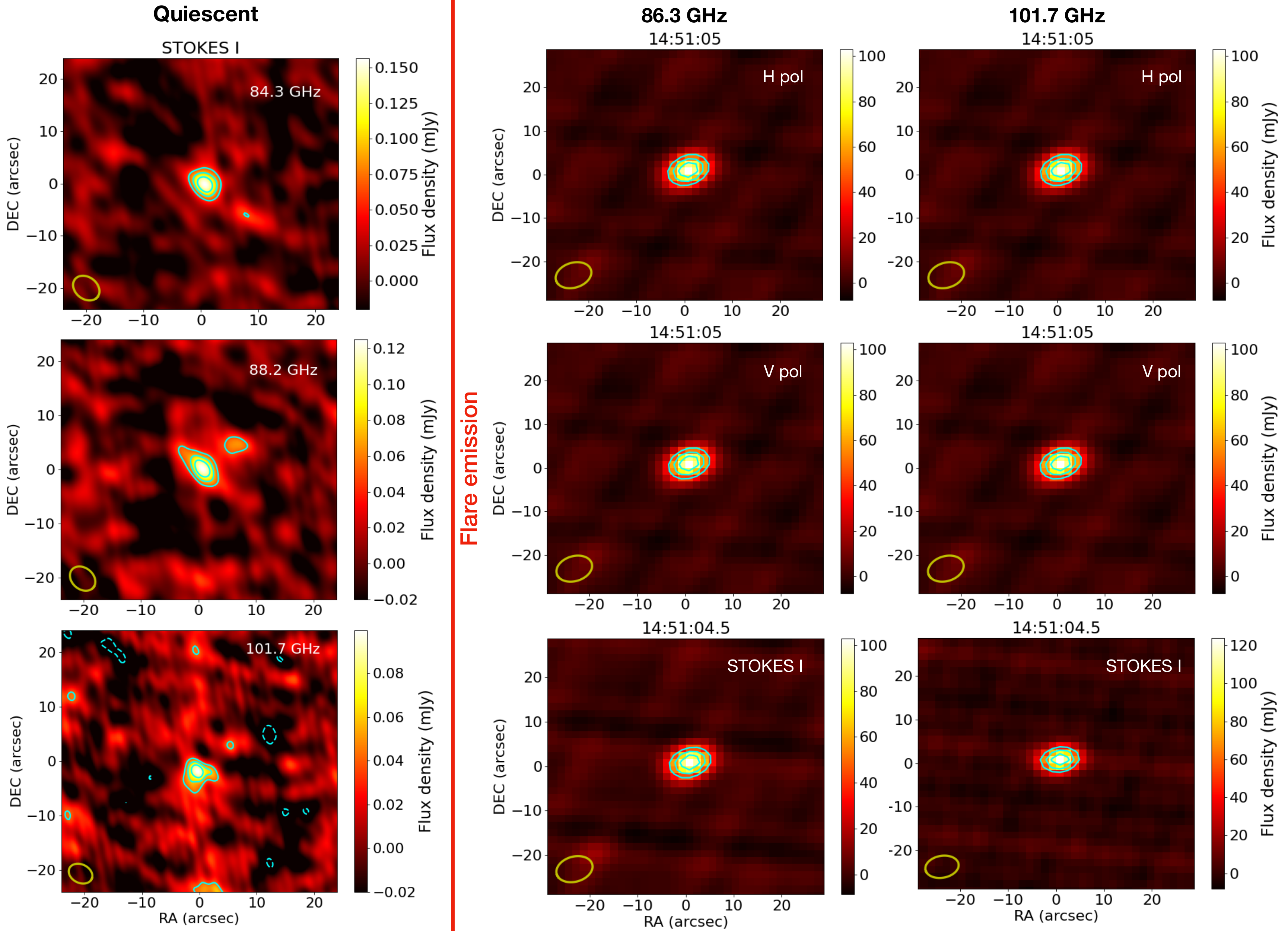}
%\vspace*{-2mm}
\caption{{ Collage of sample images. {\it Left}: Stokes I quiescent images generated by averaging the entire non-flaring period. {\it Right}: { Images during the peak flare time, 2023-07-14 14:51:05 UT, in the two polarizations and Stokes I.} Positive and negative contours are shown at -0.8, -0.6, -0.4, 0.4, 0.6, and 0.8 times the peak flux in all images. Image center: J2000 10h19m35.4s +19d52m11.1s.}}
\label{fig1:sample_imgs}
%\vspace{-0.4cm}
\end{figure*}

\begin{table}[!htb]
\centering
\begin{tabular}{|l|c|c|c|c|c|}
\hline
Integ. time\footnote{refers to time duration, between which all available unflagged data has been integrated} & $\nu$ (GHz) & ${\rm S(\nu)}$ (${\rm \mu Jy}$) & $\sigma$ (${\rm \mu Jy}$) & $\Sigma\, (\sigma)$ & $\Delta \nu$ (GHz) \\
\hline
all & 84.3 & 155 & 16 & 10 & 4.06 \\ \hline
all & 88.2 & 123 & 16 & 8 & 4.06 \\ \hline
all & 101.7 & 94 & 18 & 5 & 8.12 \\ \hline
31May & 86.2 & 135 & 21 & 6 & 8.12 \\ \hline
03Jun-06Jun & 86.2 & 137 & 17 & 8 & 8.12 \\ \hline
14Jul-19Jul & 86.2 & 134 & 27 & 5 & 8.12 \\ \hline
\end{tabular}
\caption{Quiescent Stokes I mm-\Snu. The columns (left to right) provide the time period of data integration ('all': entire non-flaring period), frequency, \Snu, RMS noise level ($\sigma$), detection significance ($\Sigma$), and bandwidth of integration.}\label{tab:mm_data}
\end{table}
\vspace{-2mm}
Table~\ref{tab:mm_data} provides the estimated quiescent mm flux density, mm-\Snu, along with the respective integration time, $\sigma$, detection significance ($\Sigma$), and the bandwidth of integration for all reliable detections. Note that $\Sigma$ mentions the signal-to-noise contrast in terms of $\sigma$. The first three rows provide the flux density obtained by averaging over the entire nonflaring period, while the last three rows concern the detections obtained by averaging over the entire LSB. Figure~\ref{fig1:sample_imgs} shows the quiescent images with full-period averaging and flare images from the peak flare epoch. The flare flux density varied within 25 to 110\,mJy, as opposed to the quiescent flux density level of $\sim$0.09–0.15\,mJy.

{Given the small fractional observing bandwidth, we derive the spectral index ($\delta$) following the common approach~\citep[e.g.,][]{Osten2006,Macgregor20_AUmicflare,Tandoi24_mmflarestars_SPT}}.
\begin{eqnarray}
\delta = {\rm \frac{\log (S_{f_l}/S_{u})}{\log(f_l/f_u)}}, \label{eqn:inx}
\end{eqnarray}
{where $S_{\rm f_l}$ and $S_{\rm f_u}$ represent \Snu\ at the lowest (f${\rm l}$) and highest (f${\rm u}$) detected frequencies. The error in the spectral index (d$\delta$) is estimated based on the limiting estimates of $\delta$ ($\delta^{+}$ and $\delta^{-}$) allowed within the flux estimation errors.
\begin{eqnarray}
\delta^{+}= {\rm \frac{\log ((S_{f_l}+dS_{f_l})/(S_{f_u}-dS_{f_u}))}{\log(f_l/f_u)}}, \label{eqn:inx+}\
\delta^{-}= {\rm \frac{\log ((S_{f_l}-dS_{f_l})/(S_{f_u}+dS_{f_u}))}{\log(f_l/f_u)}}, \label{eqn:inx-}\
d\delta = \sqrt{\frac{(\delta^{+}-\delta)^2+(\delta^{-}-\delta)^2}{2}},\label{eqn:dinx}
\end{eqnarray}
where dS${f_l}$ and dS${f_u}$ are the errors in the lower and upper band flux densities.
Given the variable frequency averaging used during the quiescent and flaring periods,
f${\rm l}$ and f${\rm u}$ differ for the two periods. When f${\rm l}$ and f${\rm u}$ are 84.2 and 101.7\,GHz respectively for the quiescent period, they are 86.3 and 101.7\,GHz respectively for the flaring period.}

\subsection{Quiescent emission}
\label{sec:sed}
\begin{figure*}[!htb]
\centering  
\includegraphics[width=0.6\textwidth,height=0.2\textheight]{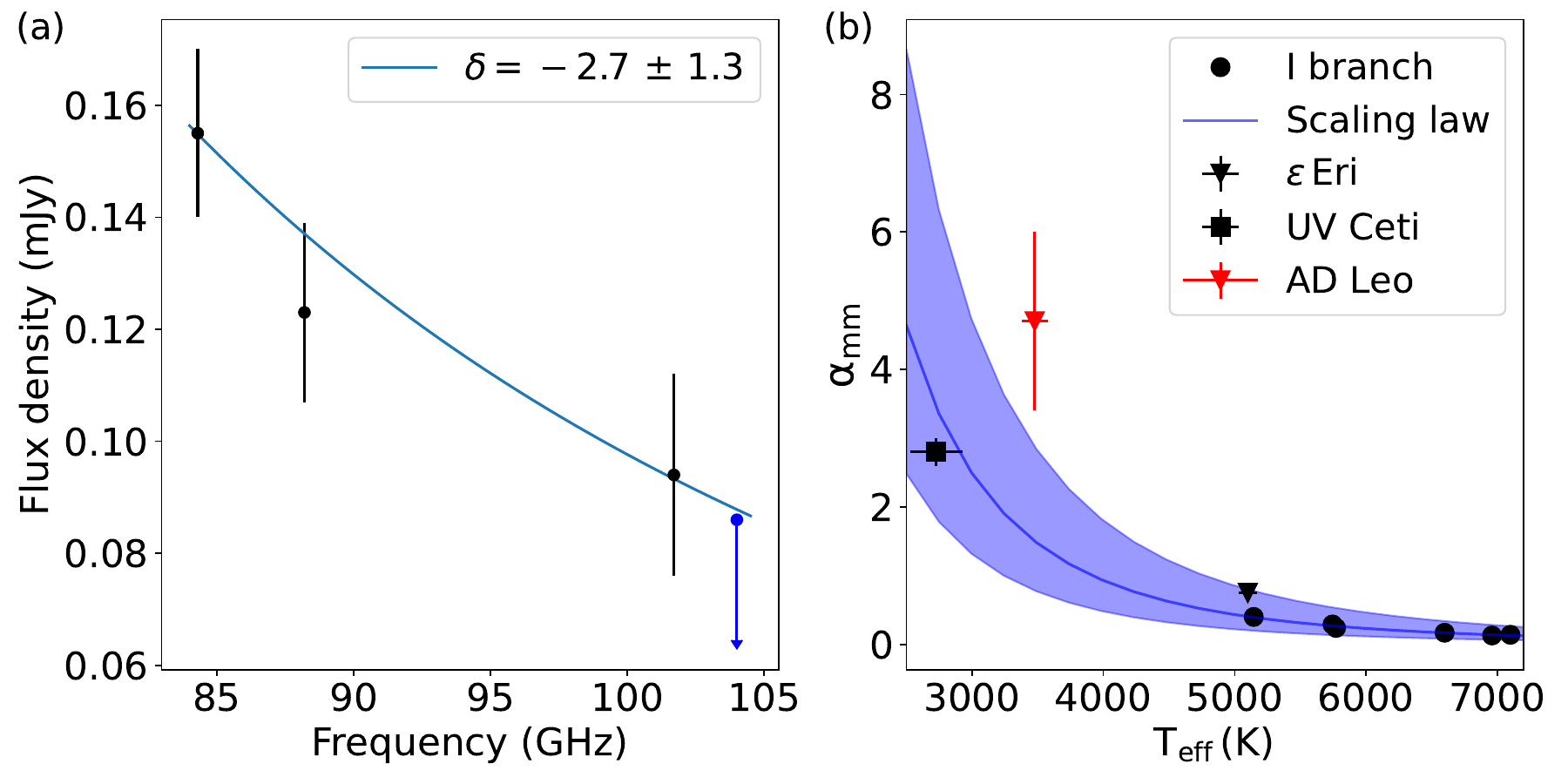}
\vspace*{-4mm}
\caption{(a): The mm-\Snu\ with the power-law model. {The flux density upper limit at 104\,GHz is marked by a downward arrow. (b): The \alfmm–\teff\ relation for the sample from \cite{Atul21_EMISSAI}, including data points for AD\,Leo and UV\,Ceti. The scaling law from \cite{Atul22_EMISSAII} is overlaid, with the shaded region representing uncertainties due to estimation errors in the model parameters.}
%{ The mm-\Snu\ with the power-law model. {Blue point with a down arrow shows the upper limit at 104\,GHz}. (b): The \alfmm\ - \teff\ plot for the sample in \cite{Atul21_EMISSAI} along with the AD\,Leo and UV\,Ceti data point. The scaling law from \cite{Atul22_EMISSAII} is shown, with the shaded region marking the error interval due to estimation uncertainties in the model parameters.}
}
\label{fig2:snu}
\vspace{-0.4cm}
\end{figure*}
The data gathered within the period from 31 May to 19 July, collectively span a full rotation phase of AD\,Leo, which has a rotation period of 2.23\,d.
{ The quiescent emission could be reliably detected in LSB at multiple epochs that correspond to different stellar rotation phases.
Since the quiescent flux density is steady across these epochs within the estimation errors (see Table~\ref{tab:mm_data}, rows 5 - 7), the spectrum obtained by combining all non-flaring data well represent the mean quiescent mm-\Snu.
Figure~\ref{fig2:snu}a shows the mm-\Snu\ of AD\,Leo. Since the three estimated flux density points are well separated from each other within estimation errors and the spectrum is steep revealing a clear variation in flux density despite the small fractional bandwidth, we performed a curve fitting using the standard routines in Python alongside with the $\delta$ estimated using Eqn.~\ref{eqn:inx}. We find that the two estimates of $\delta$ agree within the estimation errors.
The line curve in Fig.~\ref{fig2:snu} shows the {power-law} model spectrum (\Snu$\propto \nu^{\delta}$).
 Using mm-\Snu, mm-\Tbp\ and \alfmm\ were derived.
The \alfmm\ was compared with the earlier result of \cite{Atul22_EMISSAII}, to find that the value for AD\,Leo is about 3 times higher than expected from the extrapolated trend of I branch stars in the \alfmm - \teff\ plane, as seen in Fig.~\ref{fig2:snu}b.
%Simply extending this scaling law to \teff = 3477\,K, corresponding to AD\,Leo, gives an expected index of \alfmm\,=1.5, which is about 3 times lower than the observed value for AD~Leo found here.
Meanwhile, the \alfmm\ of $\epsilon$Eridani (\Prot\,$\sim$ 11.2\,d; age $\sim$ 500\,Myr; \teff$\sim$5100\,K)~\citep{Mamajek08_Ref11,Atul21_EMISSAI}, { a candidate with an activity level intermediary between C and I branch with an age $\sim$ 400 - 800\,Myr and a \Prot\,$\sim$ 11.2\,d}, also deviates from the I branch trend by a factor of 1.9.}
{Based on recent quiescent observations of UV\,Ceti (M6V; \teff: 2,728\,K, \Prot: 0.227\,d) by \cite{Plant24_UVCet_GHzQspec}, we compute an \alfmm$\sim$2.7$\pm$0.2 using the mean flux densities at 34 and 98\,GHz. This value exceeds those of all F–K dwarfs, but is lower than that of AD\,Leo.
Interestingly, the UV\,Ceti lies on the \alfmm-\teff\ scaling law of \cite{Atul22_EMISSAII} for the old sun-like stars, unlike AD\,Leo.
The age of UV\,Ceti is estimated to be approximately 1 - 5\,Gyr, based on its kinematics, position in the Galaxy, and infrared photometric data~\citep{Poveda96_UVCetitypestar_ages,kervella16_UVCeti_age_M-Rrel}.
Furthermore, UV\,Ceti exhibits significantly lower flaring rates and flare fluence across various spectral bands compared to young active dMs~\citep{Lynch17_UVCetflares154MHz,Davenport16_keplerflares_age-flaring_rel}.
Since \alfmm\ is linked to the quiescent/quasi-steady chromospheric activity~\citep{Sven16_ALMA_science,Atul22_EMISSAII}, the high \alfmm\ of AD\,Leo could indicate a significantly high heating or steady particle acceleration activity in the C branch active dM close to the fully convective internal core transition boundary~\citep{Linsky16_Stellar_chromRev}.
To explore the origin of the mm emission from active atmospheric layers, we combine it with the wide band quiescent \Snu\ and contrast it with model SEDs.}

\subsection{Comparison of \Snu\ with models}\label{sec:dis_quiescent}
{
Figure~\ref{fig3:SED_comp} compares \Snu\ with the purely photospheric model, \Smod, and \Smodc\ which features a spectroscopic data-constrained chromosphere. The bottom panels of the sub-figures show the deviation of \Snu\ from the respective models (\dsmod = \Snu/\Smod-1; \dsmodc = \Snu/\Smodc-1) %{ I guess you mean (\dsmod = \Snu/\Smod -1  ?).}
The dotted line at \dsmod\,=0 (\dsmodc\,=0) marks the perfect match condition between the model and the data.
The insets in the figure zoom in to the NOEMA observing band.
\begin{figure*}[!htb]
\centering  
\includegraphics[width=\textwidth,height=0.4\textheight]{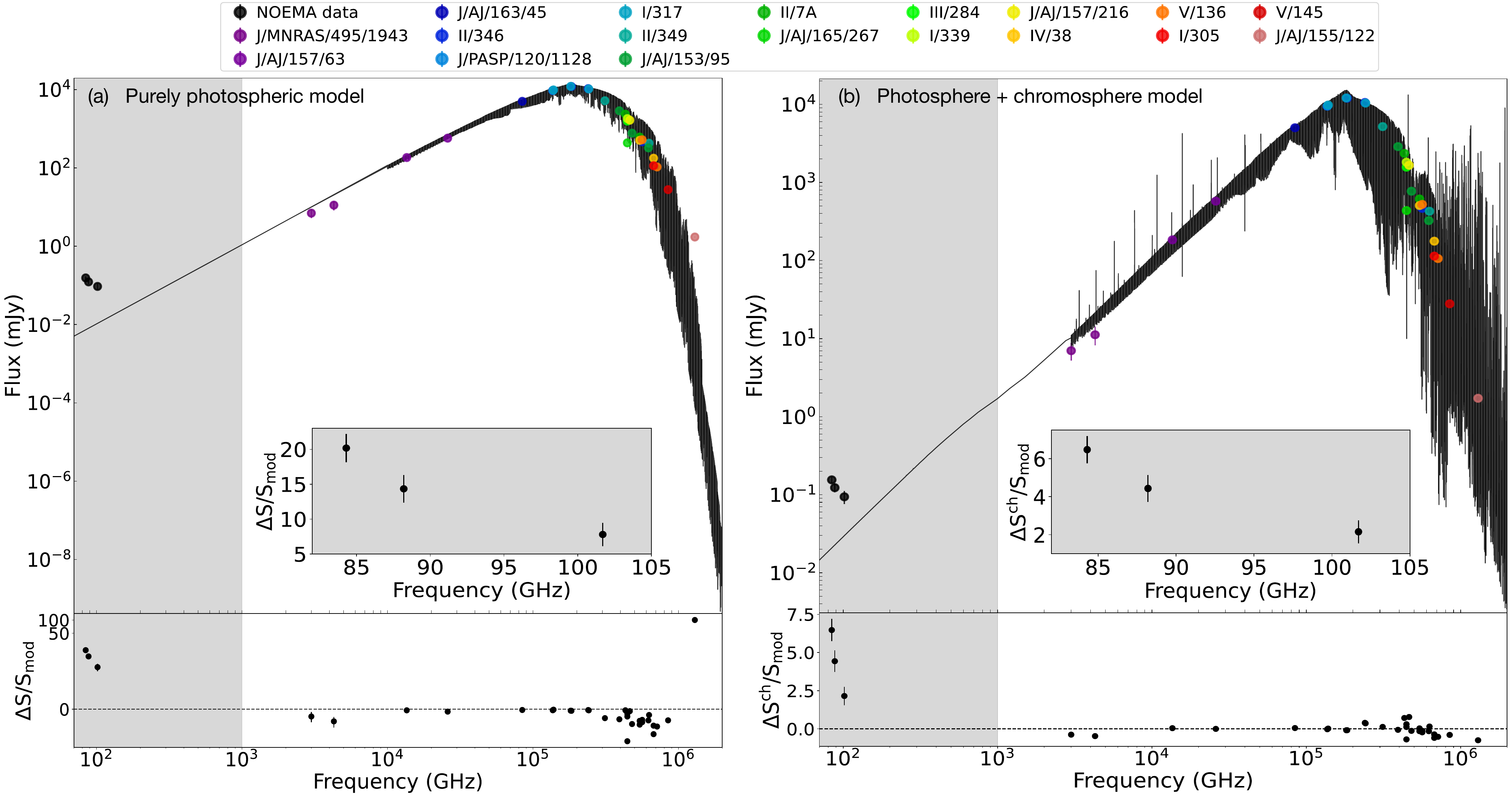}
%\vspace*{-2mm}
\caption{(a): { Comparison of SED data (\Snu) with the purely photospheric model (\Smod). The data source catalog IDs are mentioned in the legend, which can be used to retrieve them from the CDS database. The line plot shows \Smod. The {\it bottom panel} shows \dsmod\ with the dotted line marking \dsmod\ =0. Inset zooms into \dsmod\ in the mm band. (b): Same as the left panel but with the model, \Smodc. The mm-\Snu\ deviates significantly from the model SED even after incorporating a data-constrained chromosphere.}
%The power-law best-fit model is plotted (\dsmod\ $\mathrm{\propto \nu^{-\beta}}$; dotted line), and the index $\mathrm{\beta}$ is given in the legend.
}
\label{fig3:SED_comp}
\vspace{-0.4cm}
\end{figure*}
%\subsection{Model atmospheres}\label{sec:models}

\subsubsection{The purely photospheric emission model ($S_{mod}^{ph}$)}
\label{sec:photospheremodel}
A 1D local thermal equilibrium model in radiative and convective equilibrium was selected from the PHOENIX grid~\citep{husser13_PHOENIX_library,1999JCoAM.109...41H} for the physical parameters of AD\,Leo to generate a quiescent photospheric SED (\Smod). \Smod\ represents the base-level emission without upper atmospheric heating. 
%Comparing \Snu\ with \Smod\ helps assess contributions from hotter outer layers.
Figure~\ref{fig3:SED_comp}a shows that \Snu\ deviates from \Smod\ in the mm and high-energy ($\nu\gtrsim 2\times 10^5$,GHz) bands. The excess mm-\Snu\ may originate from thermal or non-thermal processes in the hot chromosphere. To investigate this, we compare the mm-\Snu\ with the UVES data-constrained thermal chromospheric model, \Smodc.
%The \dsmodc\ in Fig.~\ref{fig3:SED_comp}a shows the clear deviation of \Snu\ in the mm and high energy bands ($\nu\gtrsim 5\times 10^5$\,GHz) from \Smod. The high mm-\dsmodc\ could arise as thermal or possible non-thermal emissions from the hot chromospheric layers. To explore the origin of the excess mm emission, we first compare the mm-\Snu\ with the thermal chromospheric emission model, \Smodc, that is constrained by UVES data.

\subsubsection{The photosphere + chromosphere model (S$_{mod}^{ch}$)} \label{sec:chrom_model}
%\todo{PLEASE add model details!}
%A classical chromospheric 1D model that does not account for horizontal inhomogeneities in the atmosphere is used. Models of this type are quite successful in representing observations of various chromospheric indicator lines and continua, as was established for the Sun by the so-called VAL C model introduced by \citet{Vernazza1981}. In the dM context, models of this type have been used to compute the optical spectra based on various chromospheric emission lines in the infrared to UV regime \citep{Short1998, 2005A&A...439.1137F, Hintz2019, Peacock2019, Hintz2020}. These models typically consist of an underlying radial photospheric model in radiative equilibrium, terminated at some height to allow for a (given) temperature rise. The temperature rise can be determined semi-empirically or is parameterized, and encompasses the chromosphere and a portion of the transition region up to about 100\,000 K. The models usually do not incorporate coronal fluxes. These 1D models assume hydrostatic and ionization equilibrium since only three-dimensional models can self-consistently incorporate departures from these equilibria. However, a huge advantage of these 1D models is the capability to compute the characteristics of a larger number of lines formed in non-local thermal equilibrium (NLTE) using only modest amounts of computation time compared to three-dimensional models.
A Chromospheric 1D model typically consists of an underlying radial photospheric model in radiative equilibrium, terminated at some height to allow for a chromospheric temperature rise consistent with spectroscopic data.
We use a re-calculation of the model presented in \citet{2005A&A...439.1137F} for AD\,Leo, who also gives a detailed description of the model construction. The model was calculated using the atmosphere code PHOENIX \citep{1999JCoAM.109...41H} and the temperature structure was inferred based on the parameterized grid, that best reproduces the UVES spectrum containing various emission lines, including Balmer and \ion{Fe}{1} lines. The re-calculation used here considered all lines of \ion{H}{1}, \ion{He}{1}, \ion{He}{2}, \ion{C}{1}-\ion{C}{3}, \ion{N}{1}-\ion{N}{3}, \ion{O}{1}-\ion{O}{3}, \ion{Mg}{1}-\ion{Mg}{3}, \ion{Ca}{1}-\ion{Ca}{3}, \ion{K}{1}-\ion{K}{3}, and \ion{Fe}{1}-\ion{Fe}{3} in NLTE, compatible with the newest PHOENIX version. 
Based on this atmospheric model, \Smodc\ was computed using the radiative transfer module in PHOENIX.

 Replacing \Smod\ with the chromospheric model \Smodc\ significantly improved \Snu\ predictions above $2\times10^5$,GHz, with \dsmodc\ nearing zero. In the mm band, however, \dsmodc\ remains high ($\sim$2–6), despite being notably lower than the \Smod-based discrepancies (\dsmod\ $\sim$5–20).
The mm-\dsmodc\ also show a monotonic rise towards lower frequencies that probe higher chromospheric heights (see Fig.\ref{fig3:SED_comp}b, inset) highlighting a systematic departure from a purely thermal 1D chromospheric emission model that explains high energy spectrum well. 
The \dsmodc\ remains $>$1 even after scaling \Smodc\ assuming a 11\% filling factor for active regions with 7\,MK plasma as opposed to the 3\,MK plasma elsewhere, based on X-ray and spectro-polarimetric observations of the quiescent star\citep{Bellotti23_ZDIADLeo_Bevol, Namekata20_ADLeoXrayFlares}. 
This persistent \dsmodc\ may imply the limitations of 1D models, that cannot factor in the surface inhomogeneities. 
While emerging 3D models may improve predictions, the success of \Smodc\ in the high-energy range and the general success of 1D chromospheric models in I-branch stars~\citep[e.g.][]{2018MNRAS.481..217T, White20_MESAS, suresh20_EpsEri_RadioSEDmodel, Atul22_EMISSAII} suggest a { quasi-steady} non-thermal contribution to quiescent mm-\Snu\ in active dMs.}

\subsection{ The frequency-rising double-hump flare}
\label{sec:flare}
\begin{figure*}[t]
\centering  %\includegraphics[width=0.5\textwidth,height=0.25\textheight]{fig5_flare.pdf}
\includegraphics[width=0.7\textwidth,height=0.35\textheight]{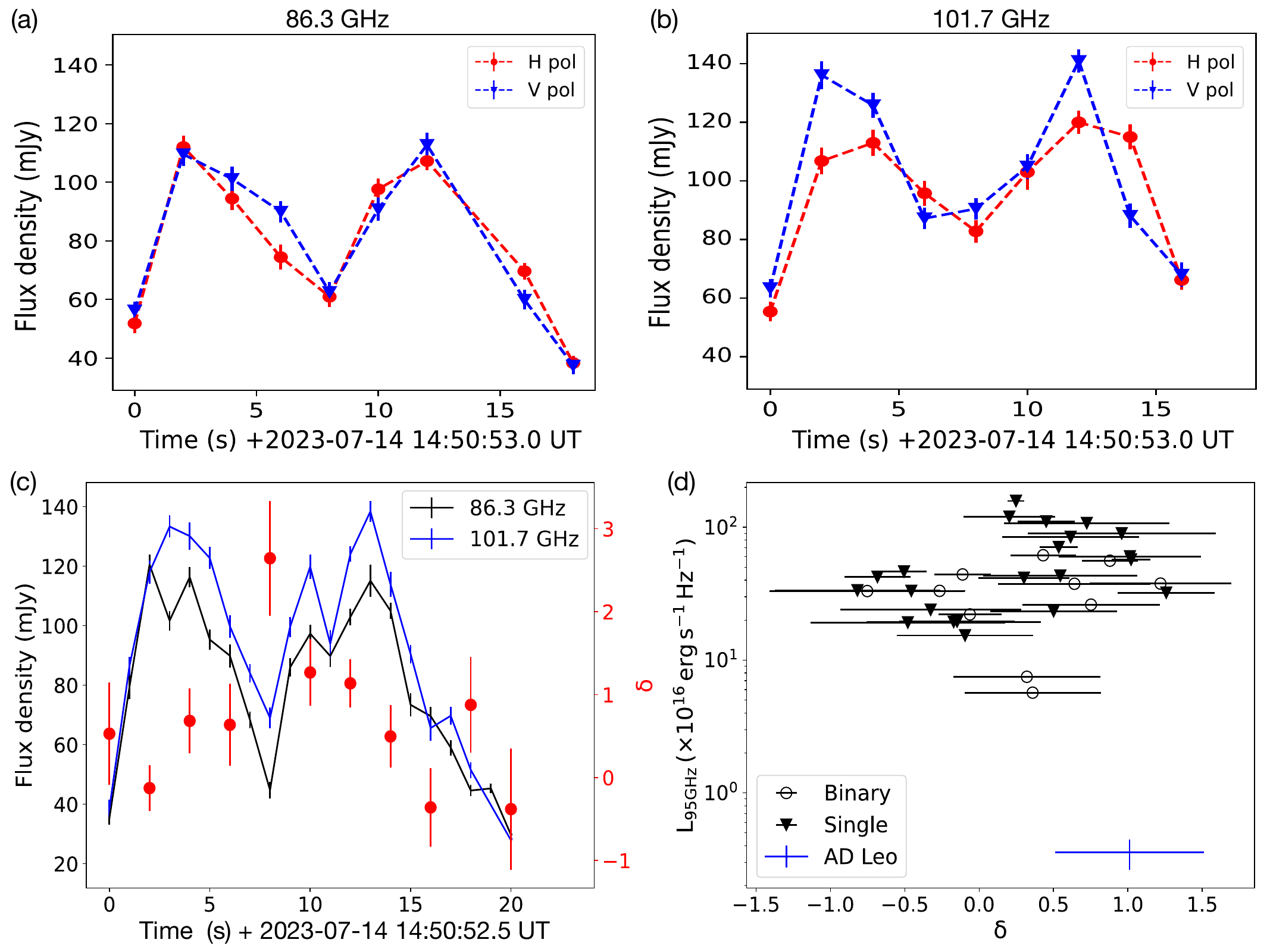}
%\vspace*{-2mm}
\caption{{Double-hump flare profile in LSB and USB.(a-b): Band-averaged H and V polarization light curves showing polarization variability. (c): Stokes I flux density and spectral index ($\delta$; in dots) evolution. (d) Comparison of the
AD Leo flare $\delta$ and L95GHz with the SPT95 sample. The potential binaries and single stars in the SPT95 sample are highlighted.
AD Leo flare has a significantly low L95GHz compared to the SPT95 sample, but a similar spectral index range.}}
\label{fig:mm_flare}
%\vspace{-0.4cm}
\end{figure*}

The complete flare flux density data and movies of the flare in the two polarization channels and Stokes I are available via Zenodo\footnote{\href{https://doi.org/10.5281/zenodo.15368248}{https://doi.org/10.5281/zenodo.15368248}} The detection significance ranges are well above 10$\sigma$ for all selected images, with an average of 24$\sigma$.
{ Figure~\ref{fig:mm_flare} show the detected double-hump flare profiles in LSB (bottom) and USB (bottom) across H and V polarizations. 
%Each hump had a width of 13\,s. 
The full-width at half maximum (FWHM) durations of the flare is {$\sim$ 18 $\pm$2\,s, combining both humps.
Due to the lack of a polarization calibration, absolute linear polarization cannot be determined.} However, since the polarization leakage between H and V channels resulting from the instrumental and atmospheric effects is not expected to vary within seconds, {the observed variability ($\lesssim$15\%) in the fractional polarization relative to the mean level (relative fractional polarization) is { most likely source-born. 
However, we emphasize the need for full Stokes calibrated data to better constrain the polarization evolution during AD\,Leo flares.} 
%We detect a relative fractional polarization variability within $\pm$15\% from the H and V light curves.
}} 
%The flare flux density is $\sim$ 2 - 3 orders of magnitude higher than the quiescent level.
%
The flare flux density ranges from $\sim$ 100 - 1000 times the quiescent level. {During periods when the flare {flux density} exceeds $\sim$ 50\% of its peak, we observe a `frequency-rising' behavior, meaning the flux density increases with frequency (Fig.~\ref{fig:mm_flare})}. Meanwhile, we could not detect a preflare source as a result of flagging about 61\% data due to bad observing conditions.

%The double-hump profile indicates a multi-epoch injection of accelerated electrons into the associated magnetic structures.

{ Recently, using mm-sky survey data from the South Pole Telescope \citep[SPT;][]{Carlstrom11_SPT}, \cite{Tandoi24_mmflarestars_SPT} published a catalog of mm-flare fluxes in 95, 150, and 220\,GHz bands for various stellar types, including dMs.
The authors released the Stokes I flare light curves and spectral indices computed using fluxes in two nearby frequencies for all detected events.
We compare our AD\,Leo flare with their 95\,GHz mm-dM flare sample, hereafter `SPT95 sample'.
%For this, Stokes I AD\,Leo flare light curves were made at 84 and 104\,GHz and $\delta$ variation was computed using Eqns.~\ref{eqn:inx} - ~\ref{eqn:dinx}.
%Unlike the quiescent emission, which has a spectral index sufficiently steep making the trend discernible given the fractional bandwidth of our observations, the flare spectrum is quite flat. So to derive a typical flare spectral index (\df) following the standard approach often used~\citep[e.g.,][]{Osten2006,MacGregor20_mmflares_dMs}, including by \cite{Tandoi24_mmflarestars_SPT}}.
%\begin{eqnarray}
%    \delta_{\rm 104}^{\rm 84} = {\rm \frac{\log (S_{f_l}/S_{f_u})}{\log(84/104)}}, \label{eqn:inx}
%\end{eqnarray}
%{ where $S_{\rm 84}$ and $S_{\rm 104}$ represents \Snu\ at 84 and 104\,GHz. The errors in \df\ are estimated by considering the furthest and the closest possible flux separations between 84 and 104\,GHz data, given the respective flux estimation error intervals.
%\cite{Tandoi24_mmflarestars_SPT} also computed the spectral index in the same way using 95 and 150\,GHz data which we will use to compare with AD\,Leo data.
{Figure.~\ref{fig:mm_flare}c} shows the Stokes I light curves and $\delta$ variation for the AD\,Leo flare computed from band-averaged LSB and USB data using Eqns.~\ref{eqn:inx} - ~\ref{eqn:dinx}. 
Unlike the quiescent emission, which has a spectral index sufficiently steep making the trend discernible given the fractional bandwidth of our observations, the flare spectrum is quite flat, {except during the dip of the double-hump profile. This variability is also detected in another set of images that we made using phase-only self-calibration with the same time–frequency averaging, confirming that the variability is source-born and not related to any synthesis imaging artifacts.}
%shows the 95\,GHz flare fluxes after scaling by distance to the star so that we can compare the flare from the data of 
%The reported 95\,GHz flux varies within $\sim$40 - 200\,mJy which a median around 100\,mJy.
Using SPT95 sample light curves, we derived the peak mm-luminosity at 95\,GHz (\lmm). The mean flare spectral index in the 95 - 150\,GHz range was taken from the SPT95 catalog.
{Figure~\ref{fig:mm_flare}d} shows the data for the SPT95 sample separating the single and potential binary dMs, and the AD\,Leo flare estimates.
The \lmm\ value interval for the AD\,Leo flare represents the peak luminosity range observed across the NOEMA sub-bands and the flux density estimation errors. Meanwhile, the \df\ value interval reflects the variation during the AD\,Leo flare, and the $\delta$ estimation errors.
The AD\,Leo flare is found to have a similar \df\ value range as the SPT95 sample, but its \lmm\ is about 1-2 orders of magnitude lower than the sample.
}

% \textcolor{red}{[Citations from EOVSA, BIMA -- please add]}. 

\section{Discussion}\label{sec:disc}

\subsection{Quiescent spectrum}
\label{sec:dis_quiescent}
{The significant deviation of \alfmm\ from the trend of I-branch stars and UV\,Ceti, and the failure of the approach to explain the mm-\Snu\ based on a 1D chromospheric emission model constrained by UVES spectroscopic data are notable. The latter approach to explain the mm-\Snu\ as thermal chromospheric emission has been quite successful in the case of I-branch stars. The high \dsmodc\ values questions the general expectation of the chromospheric thermal origin of quiescent mm emission in cool stars when it comes to active young dMs like AD\,Leo and strongly suggest the need to incorporate non-thermal emission mechanisms.}
%\subsubsection{{ Possibility of non-thermal gyrsynchrotron}}\label{sec:gyro_poss}
{The observed mm-\Snu\ index, $\delta = -2.7$, can arise via either as an optically thin non-thermal gyrosynchrotron emission or various coherent emission processes~\citep[][]{ginzburg1958,Tsytovich69,melrose1972,dulk79_gyro-thermal,Dulk85_FlareradioEmissMech,melrose2017_coherentEmissMech}.
{ A full Stokes spectrum could help} identify any coherent emission signatures, which often show high circular polarization levels~\citep[e.g.][]{Osten2006,Atul24_ADleotypeIV}.

There have been various reports of impulsive solar and stellar flares powered by various non-thermal processes in the GHz range~\citep{2020FrASS...7...57N,2021ApJ...921....9D, Gary23_imgingSpec_rev,mondal24_Xray-uwav_Flaregyro}. 
%Irrespective of the mechanism, a steady supply of non-thermal electrons is essential to power the observed spectrum.
%Since $\delta$ and p are related as $\delta = 1.22 - 0.9$p, we obtain an electron distribution power-law index of { 4.4$\pm$1.4. 
%{ Such steep indices have been reported in gyrosynchrotron emission from solar impulsive events~\citep[e.g.,][]{2020FrASS...7...57N,Gary23_imgingSpec_rev,mondal24_Xray-uwav_Flaregyro}. 
The observed quiescent AD\,Leo spectrum could arise as a superposition of frequent small-scale impulsive events supplying supra-thermal electrons to the stellar atmosphere. 
Earlier studies of AD\,Leo had revealed signatures of persistent small-scale activity in the photometric U and V bands, spectral lines and radio frequencies within 0.2 - 1.5\,GHz that probe stellar emission across chromosphere to corona~\citep[e.g.][]{1986ApJ...305..363L,crespo06_ADLeo_Spectrallineflares,2019ApJ...871..214V,dal2020flare}.
Recently, using spectral line data, \cite{Muheki20_ADLeo_flarerate_chrom} estimated a flaring rate of a few to 10 per hour for flares of energies within 10$^{29}$ - 10$^{31}$\,erg in AD\,Leo. 
Such high rates of strong flares can provide the necessary supply of energetic particles that can drive the observed emission.
Additionally, modeling the stellar quiescent X-ray emission, \cite{Namekata20_ADLeoXrayFlares} showed the presence of a steady high-energy coronal source with an effective temperature of 7\,MK, distinct from the 3\,MK plasma that prevails elsewhere.
%Besides, a coherent burst emission duty cycle of $\sim$27\% had been estimated at 1\,GHz by \cite{2019ApJ...871..214V}. 
% We need multi-waveband follow-up studies with mm to radio spectral coverage, along with spectroscopic observations to better constrain the emission mechanism.  
Besides, spectro-polarimetric observations of quiescent AD\,Leo reveal a mean line of sight magnetic field strength of $\sim$3.5\,kG in Stokes I data~\citep{Reiners22_Bavg_dMs, Bellotti23_ZDIADLeo_Bevol}. The stellar active regions could have even higher field strengths at smaller scales in particle acceleration regions powering weak flares (energy $<$10$^{28}
$\,erg) at rates of a few every minute given the high flaring rates~\citep{Muheki20_ADLeo_flarerate_chrom}.
All the aforementioned aspects point to a likely steady supply of accelerated electrons in the atmosphere that could power the observed quiescent non-thermal emission.
Simultaneous soft X-ray, radio and spectral line observations are essential alongside wide band full Stokes mm data to better constrain the stellar quiescent emission model.}

\subsection{{The frequency-rising double-hump flare}}
%As mentioned in Sec.~\ref{sec:flare} we report a 26\,s long double hump flare that caused a 2 to 3 orders of magnitude rise in mm-\Snu\ across the 84 - 104\,GHz (see, Fig.~\ref{fig:mm_flare}). 
{We report the detection of a second-scale double-hump frequency-rising mm flare in 85 - 105\,GHz range for possibly the first time in an active dM. 
The flare ${\rm T_B}$ reaches an order of 10$^7/f\,$K, where $f$ is the fractional size of the flare region with respect to the stellar disk. 
Modeling the modulations observed in various optical flares in AD\,Leo,
a few recent studies estimated the typical size of the flaring region~\citep[e.g.][]{2022A&A...667L...9S,ram25_ADLeoflares_multiwb}.
Based on these results, we assume f $\sim$ 1\%  implying a ${\rm T_B}\sim 10^9\,$K for our mm-flare.
Such high ${\rm T_B}$, along with the second-scale variability in the Stokes I spectral index that occasionally exceeds 2 is difficult to explain
via a thermal flare emission mechanism~\citep[][]{Fleishmann10_newMmSolFlareMech,Sven16_ALMA_science,2020FrASS...7...57N,Fleishman21_Gyro_F-F_EmissCalc}.
%However thermal signatures are expected to have a much larger duration than what is observed here since the cooling timescale for a flare that produces 2 - 3 orders of magnitude energy rise is expected to be at least several minutes to hours~\citep[e.g.,][]{RTV1978,crespo06_ADLeo_Spectrallineflares,kowalski24_stellarflare_rev}.
%The variability in the relative fractional linear polarization also suggest a non-thermal emission process, since thermal emission is expected to be unpolarized. %Instrumental effects are not expected to produce such polarization variability despite clearly detecting the source images beyond 4 - 5 $\sigma$ detection significance during the flare.
%Faraday conversion due to propagation through the ambient medium could generate linear polarization, but the observed polarization variability in a span of a few seconds from positive to negative is not reproducible by any physically reasonable variations or the instrument. 
%The second-scale variability in the Stokes I flux, with varying mostly positive \df\ values which at an instant tends to exceed 2, is also difficult to obtain with thermal processes from a flare source especially given the high source ${\rm T_B}$ (see, Fig.~\ref{fig:mm_flare}c)~\citep{Fleishmann10_newMmSolFlareMech}.
%This is because to have a spectral index above 2, one would require absorption along the line of sight from denser plasma which would subsequently decrease the total observed intensity, making the observed high ${\rm T_B}$ difficult to explain within a thermal emission framework~\citep{Fleishmann10_newMmSolFlareMech}.
Various coherent and incoherent non-thermal emission mechanisms modulated by local or line of sight dynamics of plasma and magnetic fields have been proposed to explain the mm-flares with second-scale modulations in $\delta$ and polarization~\citep[e.g.,][]{silva1997,2001ApJ...548L..95K,trottet08_2.2MeVline+200GHzsrc,Fleishmann10_newMmSolFlareMech,zaitsev14_plasmaemissMod_Mmpulses,2014ApJ...791...31K}.
Most of these models are proposed in the context of solar flares, and local physical parameters can differ between stars, but plasma physics and emission mechanisms are extendable to cool stars~\citep{guedel02_Rev_stellarRadioEmiss,2004NewAR..48.1319W,Atul23_mmRadioTomo}.
The solar mm-flares for which a mechanism could be identified with reliable conviction had the advantage of simultaneous multi-waveband data, wide band GHz spectra with polarization calibration, and spatially resolved imaging~\citep[e.g.,][]{kaufmann85_95GHzsolarflare,Asai01_17GHzQPPs,Shen23_mmQPP_Xclassflare}.
%Hence, we refrain from making strong conclusions using our mm-flare data and emphasize the need for future simultaneous multi-waveband spectro-polarimetric observations in X-ray to GHz bands.
%Hence, to conclude on the exact nature of the non-thermal emission mechanism in the stellar context, we need wide spectral coverage, full Stokes polarization to infer the absolute circular and linear polarization variability, and multi-waveband spectro-polarimetric observations.
The multi-waveband spectro-polarimetric data is vital to probe the thermal and non-thermal dynamics across various atmospheric heights, making up for the lack of spatially resolved stellar observations.
}

%{ The seconds-scale variability in Stokes I flux density with significant relative fractional polarization and spectral index variability are all features previously identified in microwave gyrosynchrotron flares in sun and stars~\citep[e.g.,][]{Dulk85_FlareradioEmissMech,2020FrASS...7...57N}.}
{Meanwhile observation of the second-scale modulations in the Stokes I flare profile along with the frequency-rising character is noteworthy. 
Several cases of $\lesssim$1\,s scale quasi-periodic modulations in flare profiles have been reported in solar flares across 90 - 500\,GHz, unlike in stellar flare research~\citep[see,][for an overview]{Krucker13_mmSolarflareRev}.
These are often signatures of multiple episodes of particle acceleration and injection into the local magnetic field structures, irrespective of the details of the emission mechanism~\citep[e.g.,][]{silva1997,raulin03_fastmmQPPs_sun}.
As mentioned earlier, without full Stokes spectra we cannot distinguish between a coherent and incoherent emission process, both of which can produce all observed emission features in a disk averaged stellar flare spectrum.
%The solar flares in the microwave band usually have a spectral turnover below $\sim$20\,GHz, and show a trend of decreasing flux with increasing frequency in the mm band~\citep{2020FrASS...7...57N,Gary23_imgingSpec_rev}. 
%The turnover frequency is a function of \nuB, which is proportional to $\gamma^2$B, where $\gamma$ is the Lorentz factor and B is the magnetic field strength. 
%Given that AD\,Leo has 10 times stronger active region magnetic fields~\citep{Reiners22_Bavg_dMs,Bellotti23_ZDIADLeo_Bevol} which could accelerate electrons to a few $\gamma$ during strong flares~\citep{Stepanov01_micrwaveflare_ADLeo5GHz,2022A&A...667L...9S}, the turnover frequency can be extended beyond 100\,GHz resulting in the observed frequency-rising spectrum. 
%A wider band  multi-frequency mm observation is required to estimate the quiescent and flaring spectral profiles.
}

{Although the \lmm\ and duration of the AD\,Leo flare lie within the ranges reported for 230\,GHz flares in AU\,Mic~\citep{Macgregor20_AUmicflare} and Proxima\,Cen~\citep{MacGregor18_proxima_flares,Howard22_ProxCenFlare}, the AD\,Leo flare stands out by exhibiting a frequency-rising behavior when the flare flux density is above 50\% of the peak, unlike the others.
Together the flares in AU\,Mic, Proxima\,Cen and AD\,Leo span a luminosity range of $\sim$10$^{13}$ - $5\times $10$^{15}$\,erg\,s$^{-1}$\,Hz$^{-1}$ and mean spectral index rage within -2 to 1.5.
Meanwhile, these individual events are 1 - 3 orders of magnitude weaker than dM flares (luminosity range: $\sim 5\times$ 10$^{16}$ - 10$^{18}$\,erg\,s$^{-1}$\,Hz$^{-1}$) in the SPT95 sample with a similar \df\ range (see, Fig.~\ref{fig:mm_flare}d), suggesting that the SPT95 flares are stronger cousins of the flares reported in AU Mic, Proxima Cen, and now in AD\,Leo.}
{Given the SPT sensitivity and the survey cadence of $\sim$2\,minute, the survey will miss relatively weak flares with sub-minute scale durations.} 

\section{Conclusion}\label{sec:conclusion}
The quiescent millimeter emission from cool main sequence stars (F - M type) is thought to primarily originate as chromospheric thermal emission across height~\citep[e.g.,][]{Sven16_ALMA_science,2018MNRAS.481..217T,White20_MESAS,Atul21_EMISSAI}. Hence \cite{Atul22_EMISSAII} proposed to use the mm brightness temperature (\Tbp$\sim{\rm \nu^{-\alpha_{mm}}}$) spectral index, \alfmm\, as a proxy to chromospheric thermal gradient and derived its trend with the effective surface temperature, \teff, for old ($>$1\,Gyr) slow-rotating sun-like (`I' branch) stars. 
%Though their analysis suffered from the scarcity of data, it demonstrated the strength of mm-quiescent spectral characterization to constrain the stellar atmospheric models and explore the nature and emergence of stellar activity across the main-sequence. 
%The stars form two activity populations `C' and `I' in the \Prot\-age-\teff\ plane, with the former being young ($<$1\,Gyr) and fast rotating (\Prot$<$5\,d) the latter being vice-versa. 
{ We present observations of quiescent emission from a young, rapidly rotating dM, AD\,Leo, in the millimeter band (mm-\Snu). The mm-\Snu\ of AD\,Leo is $\sim$ 2 - 7 times brighter than the thermal spectrum (\Smodc) expected from a 1D atmospheric model that includes the chromospheric component and is constrained by optical and UV spectroscopic data.}
The \alfmm\ of AD\,Leo is also significantly higher than the values observed in {$\epsilon$Eridani and the old dM, UV\,Ceti.}
The deviation of mm-\Snu\ from \Smodc\ and the high \alfmm\ suggest {the relevance of non-thermal processes from supra-thermal electrons, in the quasi-steady mm-emission.
Meanwhile, the \alfmm\ of UV\,Ceti is consistent with the prediction from the scaling law for old sun-like stars.
%An optically-thin non-thermal gyrosynchrotron emission { or a superposition of coherent impulsive bursts can explain the mm-\Snu.
} 

{We report an $\sim$18\,s long double-hump flare with a `frequency-rising' nature, meaning the flux density rises with frequency, during times when the flare flux density was $\gtrsim$50\% of its peak. To the best of our knowledge, a frequency-rising double-hump flare with second-scale spectro-temporal variability is not reported in dMs, though several reports exist in the Sun. 
The flare brightness varies by $\sim$ 200 - 1000 times the quiescent level in a few second timescale with significant variability in the Stokes I spectral index.
%The emission is powered by a non-thermal mechanism. 
%Given the order of magnitude larger magnetic fields in AD\,Leo capable of accelerating particles to a mildly relativistic regime, it is possible to produce the observed high ${\rm T_B}$, highly variable frequency-rising non-thermal flare.
The double-hump flare profile suggests multiple particle acceleration episodes producing non-thermal emission.
The AD\,Leo flare is $\sim$ 1–3 orders of magnitude fainter than the 95\,GHz dM flares observed by SPT \cite{Tandoi24_mmflarestars_SPT}, although its spectral index (\df) remains within the SPT flare range.
%But the spectral index falls well within the range of SPT flares.
However, the luminosity and duration of the flares reported in Proxima\,Cen and AU\,Mic compare well with the AD\,Leo flare, making these individual events weaker cousins of SPT flares.
In order to better constrain physical mechanisms contributing to the quiescent and flare emission in the mm bands we need simultaneous spectro-polarimetric observations across mm to X-ray wavelengths.}

\begin{acknowledgements}
AM is supported in part by NASA's STEREO project and LWS program. AM thanks Dr. Peter Young for the generous funds to support the publication.
This work is partly supported by the Research Council of Norway through the EMISSA project (project number 286853) and the Centres of Excellence scheme, project number 262622 (``Rosseland Centre for Solar Physics''). This work is based on observations carried out under project number W22BF with the IRAM NOEMA Interferometer [30m telescope]. IRAM is supported by INSU/CNRS (France), MPG (Germany) and IGN (Spain). AM thanks the
staff of the NOEMA, especially Vinodiran Arumugam, for enabling the observations and the support in data calibration. 
This research made use of NASA's Astrophysics Data System (ADS). 
AM acknowledges the developers of the various Python modules namely Numpy, Astropy and Matplotlib. AM also thanks the developers of CASA and CLIC.
\end{acknowledgements}
\facilities{NOEMA}


\software{Numpy~\citep{numpy}, Astropy~\citep{astropy}, Matplotlib~\citep{matplotlib}, Pandas~\citep{reback20_pandas},
Scipy~\citep{scipy}, 
CASA, CLIC}

\bibliographystyle{aasjournal}
\bibliography{paper} 
\end{document}